\documentclass[aps,prl,twocolumn,floatfix,amsmath,amssymb]{revtex4}
\usepackage[final]{graphicx}
\usepackage{bm}
\usepackage{afterpage}
\usepackage{color}
\usepackage{comment}
\usepackage[colorlinks=true,urlcolor=blue,linkcolor=blue,citecolor=blue]{hyperref}

\emergencystretch=\maxdimen
\begin{document}

\title{Quantum magnetism of topologically-designed graphene nanoribbons}

\author{Xingchuan Zhu$^{1}$}
\author{Huaiming Guo$^{2}$}
\email{hmguo@buaa.edu.cn}
\author{Shiping Feng$^{1}$}
\affiliation{$^1$Department of Physics, Beijing Normal University, Beijing, 100875, China}
\affiliation{$^2$Department of Physics, Key Laboratory of Micro-Nano Measurement-Manipulation and Physics (Ministry of Education), Beihang University, Beijing, 100191, China}

\begin{abstract}
Based on the Hubbard models, quantum magnetism of topologically-designed graphene nanoribbons (GNRs) is studied using exact numerical simulations. We first study a two-band Hubbard model describing the low-energy topological bands using density matrix renormalization group (DMRG) and determinant quantum Monte Carlo (DQMC) methods. It is found the spin correlations decay quickly with the distance, and the local moment is extrapolated to zero in the presence of symmetry-breaking terms. The results show that the two-band Hubbard chain is nonmagnetic, which is in contrast to the mean-field calculation predicting a critical interaction for the magnetic transition. We then include the Hubbard interaction to the topological-designed GNRs. For large interactions, the spin correlations keep finite for all distances, and the magnetic order develops. The local moment is extrapolated to almost zero for weak interactions, and begins to increase rapidly from a critical interaction. The estimated critical value is much larger than the realistic value in graphene, and we conclude the experimentally relevant GNRs is nonmagnetic, which is consistent with the experimental results.

\end{abstract}


\maketitle


\subsection{Introduction}
Graphene nanoribbon (GNR) is a planer strip of graphene with extraordinary electronic and physical properties\cite{yazyev2010,son2006b,yazyev2013}. It is first introduced as a theoretical model to study the size effect in graphene\cite{nakada1996}. Recent advance in bottom-up techniques have allowed production of atomically precise GNRs with armchair, zigzag and other sophistated edges\cite{kimouche2015,ruffieux2016}.
GNRs have potential applications in the next-generation nanoelectronics, and have been extensively investigated.

GNRs exhibit electronic properties that are not present in graphene\cite{ruffieux2016}. In the simplest form, the edges can be either armchair or zigzag. Tight binding calculations predict that zigzag GNRs are always metallic,
while armchair ones can be insulating or metallic. The graphene features two inequivalent Dirac points, which are characterized by $\pm \pi$ Berry phases\cite{castro2009}. In the presence of zigzag edges, the two Dirac points are projected to different momenta of the one-dimensional (1D) Brillouin zone, and flat bands connecting them appear. The low-energy states on the flat bands are localized at the edge. The large density of states at the Fermi energy is sensitive to the electron-electron interactions, and the nanoribbon has an instability to magnetic ordering. Remarkable edge magnetism develops in the weak-coupling regime $U/t<2$, where bulk magnetic order is absent\cite{sorella1992,paiva2005,meng2010,sorella2012}. The magnetic moments form long-ranged ferromagnetic order along the zigzag edge while they are antialigned at the opposite edge\cite{feldner2011,hikihara2003,feldner2010,luitz2011,schmidt2010,wessel2013,hagrmasi2016}. The opposite spin polarization along the zigzag edges opens a band gap, and the system becomes a semiconductor. Recent experiments provide direct evidence of the edge magnetism\cite{joly2010,magda2014}. In particular, it is found that the magnetic order on zigzag edges can be stable even at room temperature\cite{magda2014}.

The gap values of the armchair GNRs depend on width. When the number of dimer lines across the ribbon is $N=3p+2$ with $p$ a positive integer, the armchair GNRs are metallic. Otherwise, they are insulating with the gaps inversely proportional to the widths\cite{son2006,brey2006,ezawa2006}. Recently, the topological properties of the insulating armchair GNRs are investigated, which demonstrates that different widths and end terminations lead to distinct topological phases\cite{cao2017,lin2018}. The topology is characterized by a $\mathbb{Z}_2$ invariant, and is manifested by localized boundary state between two segments with different topologies. Later experiments confirming the predictions are reported\cite{groning2018,rizzo2018}. Furthermore, the junction states are successfully used as building block to engineer the famous Su-Schrieff-Heeger topological model\cite{groning2018,rizzo2018,su1979}.

The armchair GNRs have no edge states, and the critical interaction to antiferromagnetism is similar to that of bulk graphene. In the topologically-designed GNRs, there are zigzag-edge segments at the boundary, which suggests the possibility of magnetic ordering at weak interactions. While the mean-field (MF) calculation finds a critical interaction smaller than that of bulk system ($U_c=2.23t$)\cite{sorella1992}, it is still larger than the realistic value in graphene, and supports that the experimentally realized structures are nonmagnetic. Since the quantum fluctuation is strong in two dimensions, it is very necessary to perform an exact numerical study of the effect of interactions, and validate the absence of the magnetism.

In the manuscript, we study quantum magnetism of topologically-designed GNRs based on the Hubbard models using exact numerical simulations. We first derive a two-band model for the low-energy topological bands, and study the corresponding 1D Hubbard model using DMRG and DQMC methods. It is found the spin correlations decay quickly with the distance, and the local moment is extrapolated to zero in the presence of symmetry-breaking terms. The results show that the two-band Hubbard chain is nonmagnetic. We then include the Hubbard interaction to the topological-designed GNRs, and calculate the spin correlations and the local moments in the presence of symmetry-breaking terms. We find the magnetic order develops from a critical interaction, which is much larger than the realistic value in graphene. We conclude the experimentally relevant GNRs is nonmagnetic, which is consistent with the experimental results.

\begin{figure}[htbp]
\centering \includegraphics[width=8cm]{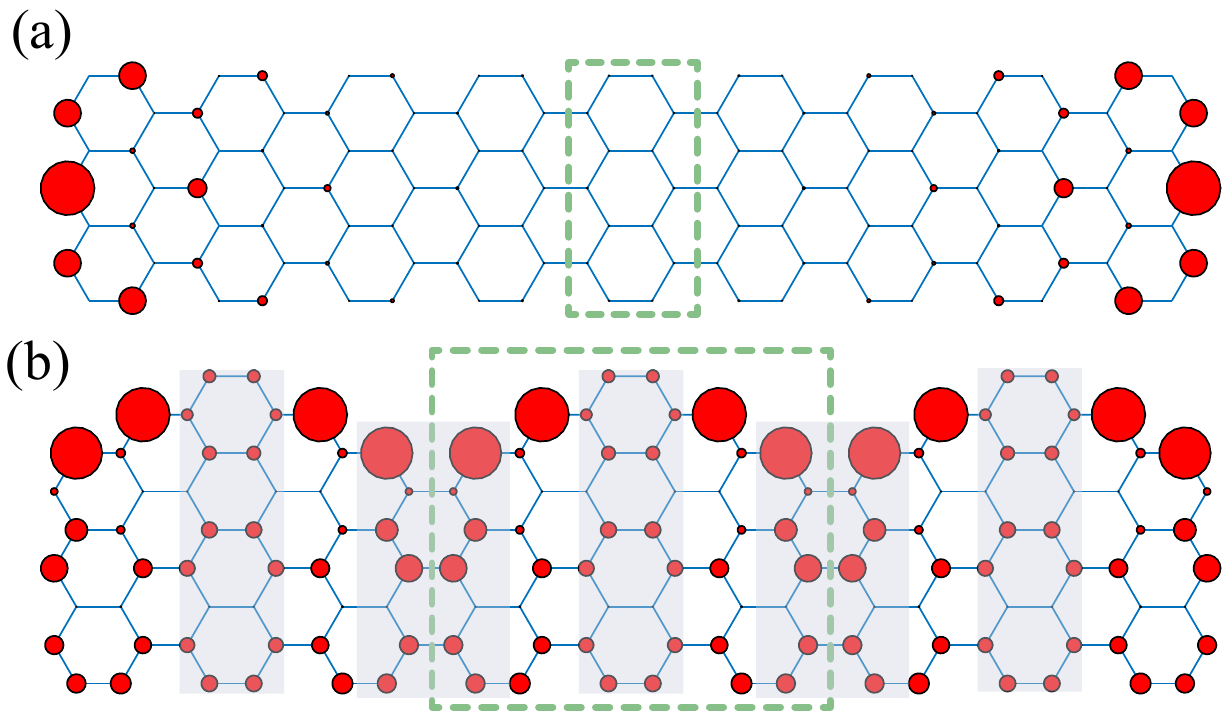} \caption{(a) Schematic structure of a pristine 7-AGNR. (b) The geometry of a topologically-designed GNR based on the pristine 7-AGNR. In (a) the pristine 7-AGNR with a zigzag termination is a 1D topological insulator, and the wavefunction of the zero boundary mode is plotted (red filled circles with the radius denoting the wavefunction amplitude). In (b) we also show the wavefunction of the designed topological band, which is mainly distributed on the two outmost sites of each zigzag shoulder. The unit cell is enclosed by green dashed lines, which have $14 (48)$ sites in (a)[(b)].}
\label{fig1}
\end{figure}

\subsection{The model and method }
We consider a nearest-neighbor (NN) tight-binding Hamiltonian on a topologically-designed GNR with the geometry shown in Fig.~1,
\begin{equation}\label{eq1}
H_0=-\sum_{\langle lj \rangle \sigma}
t c^\dag_{j\sigma}c^{\phantom{\dag}}_{l\sigma},
\end{equation}
where $c_{j\sigma}$ and
$c^{\dag}_{j\sigma}$
are the annihilation and creation operators at site $j$
with spin $\sigma=\uparrow, \downarrow$.
The hopping amplitudes between the NN sites
$l$ and $j$ are $t$, which is about $3~eV$ in graphene.

The geometry in Fig.~1(b) is obtained by adding small segments with zigzag edges periodically on the pristine armchair GNR with the width $N=7$ ($7$-AGNR). The resulting geometry is a superlattice of alternating unit cells of $7$-AGNR and $9$-AGNR. The topological property of the $7(9)$-AGNR composed of such unit cells [see shaded region in Fig.~\ref{fig1}(b)] is described by a $Z_2$ invariant: $Z_2=\frac{1\pm(-1)^{ \lfloor \frac{W}{3}\rfloor+\lfloor\frac{W+1}{2}\rfloor }}{2}$, where $\lfloor x\rfloor$ is the floor function and the sign is negative (positive) for $W=7(9)$\cite{cao2017}. The $Z_2$ value is $0(1)$ for $7(9)$-AGNR, and the topological and trivial unit cells alternate along the chain. Thus the localized boundary states between adjacent unit cells may form low-energy dispersing bands in the gap.

Performing Fourier transformation, the Hamiltonian in Eq.(\ref{eq1}) becomes $H_{0}=\sum_{\bf k}\psi^{\dagger}_{\bf k}{\cal H}({\bf k})\psi_{\bf k}$, where the basis is $\psi_{\bf k}=(c_{1,{\bf k}}, c_{2,{\bf k}},..., c_{N_{s},{\bf k}})^{T}$ ($N_s$ is the number of sites in a unit cell), and ${\cal H}({\bf k})$ is the $N_s$-by-$N_s$ Hamiltonian matrix in the momentum space. By diagonalizing ${\cal H}({\bf k})$, the band structures are directly obtained, and are shown in Fig.~2. Figure~2(a) shows the band structure of the pristine $7$-AGNR for comparison, which has been known to be a semiconductor with the gap size about $0.5t$. By including small segments, two new dispersing bands appear in the gap[see Fig.~2(b)], and their wave functions mainly distribute on the zigzag sites of the small segments.

Since the two low-energy bands in Fig.~2(b) are isolated from other bands, an effective model for them can be constructed using maximally localized Wannier functions\cite{marzari2012}, which is a 1D chain containing hoppings of different ranges. The main contribution is from the NN ones, and the corresponding hopping amplitudes are $t_1=0.1493t$ and $t_2=-0.1824t$, respectively. It is found that the low-energy two bands are fitted quite well only using the above NN hoppings (see Fig.~2). Thus the designed GNR realizes an effective 1D tight-binding model analog to the famous Su-Schrieffer-Heeger one.

\begin{figure}[htbp]
\centering \includegraphics[width=8cm]{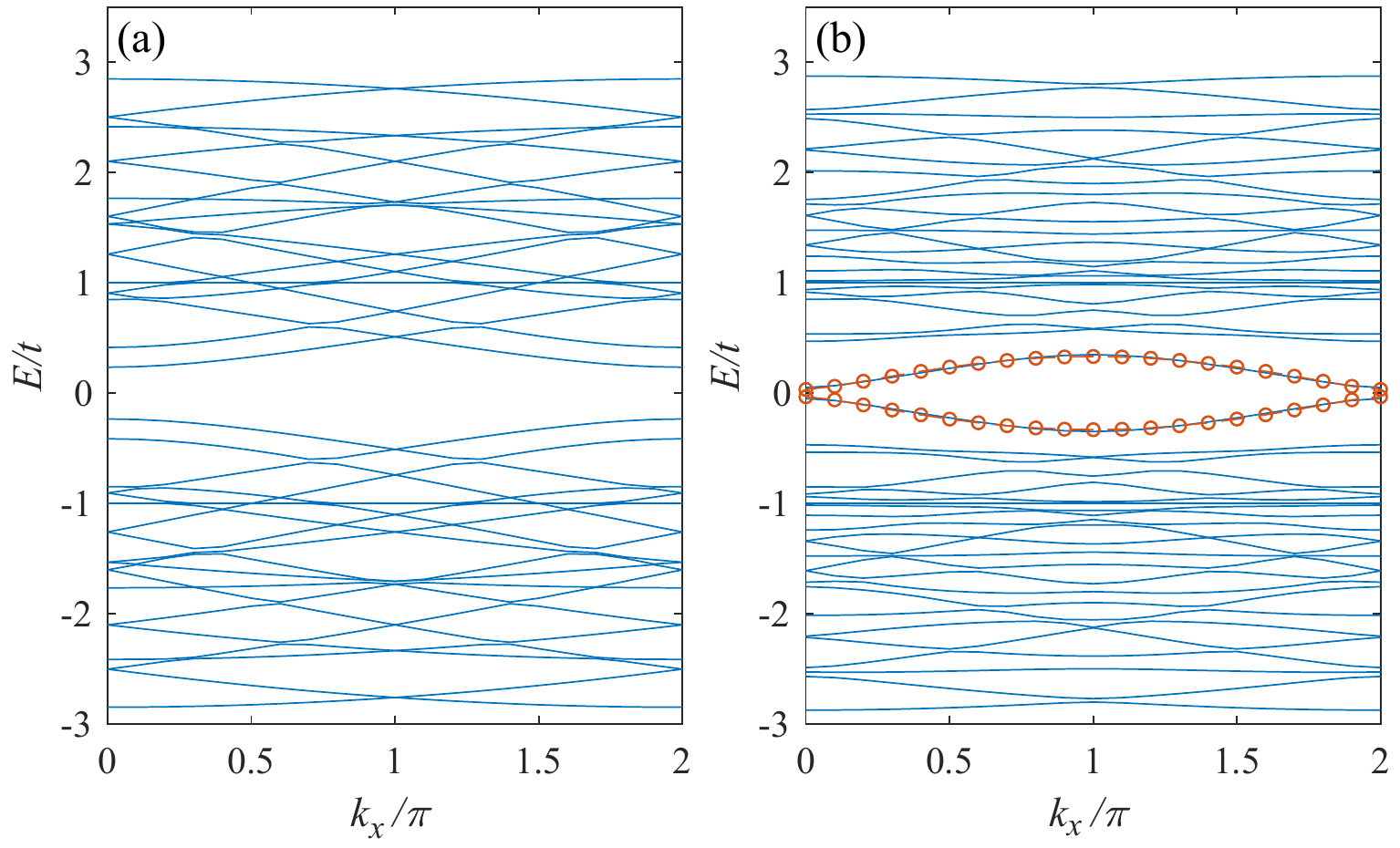} \caption{The band structures for (a) the pristine 7-AGNR, and (b) the topologically-designed GNR based on the pristine 7-AGNR. In (b), the two bands near the Fermi energy are well fitted using a 1D chain with alternating hopping amplitudes $t_1=0.1493t$ and $t_2=-0.1824t$ (orange open circles).}
\label{fig2}
\end{figure}

It has been known that the pristine $7$-AGNR supports a topological phase. The topological invariant depends on the shape of its termination, and the zigzag termination yields a nontrivial topological invariant. It is also desirable to know the topological class of the above topologically-designed GNR. The 1D topological property is characterized by the Berry phase\cite{berry1,berry2,wang2014,resta1994,xiao2010},
\begin{eqnarray}\label{eq2}
 \gamma=i\oint \langle \psi_{k}|\frac{d}{d k}|\psi_{k}\rangle d k,
\end{eqnarray}
where $k$ varying from $0$ to $2\pi$ is the wave vector, and $\psi_{k}$ is the periodic part of the Bloch wave function corresponding to $k$.
The Berry phase of the Hamiltonian Eq.(\ref{eq1}) on the geometry shown in Fig.~2 has a nontrivial value $\pi$, corresponding to which a pair of zero boundary modes appear under open boundary condition.

To study the magnetic property of the topologically-designed GNR, we consider the Hubbard interaction, which writes as,
\begin{equation}\label{eq3}
H_{U}=\sum_{i}
U(n_{i\uparrow}-\frac{1}{2})(n_{i\downarrow}-\frac{1}{2}),
\end{equation}
where $n_{i\sigma}=c^{\dagger}_{i\sigma}c_{i\sigma}$ is the number operator of fermion and $U$ is the strength of the on-site interaction. In the following discussions, we include the Hubbard term into both the Hamiltonian in Eq.(1) and the two-band effective model, and study the induced magnetism.

The interacting Hamiltonian is solved numerically by means of the
DQMC and DMRG methods\cite{white89,dqmc,white92}. In the DQMC approach, one decouples the on-site
interaction term through the introduction of an auxiliary
Hubbard-Stratonovich field (HSF).  The fermions are integrated
out analytically, and then the integral over the HSF is performed
stochastically.  The
only errors are those associated with the statistical sampling, the
finite spatial lattice
and inverse temperature discretization.
All
are well-controlled in the sense that they can be systematically reduced
as needed, and further eliminated by appropriate extrapolations. The DMRG method is most efficient for the ground state of 1D systems\cite{Bauer2011}. We use it to solve the effective model directly at zero temperature. In the following calculations, we focus on half-filled bands, when the DQMC method is free of the infamous `minus-sign problem'.

To study the magnetic behavior, we measure the local moment $m^z_j=\langle S^z_{j}\rangle$ with $S^z_{j}=\frac{1}{2}(n_{j\uparrow}-n_{j\downarrow})$. Since the original Hamiltonian preserves the $SU(2)$ symmetry, a symmetry-breaking term should be included to induce the magnetism. We break the symmetry of the magnetic phase to $z-$axis by adding an alternating Zeeman term $H_{B}=\sum_{j} B_{j} (n_{j\uparrow}-n_{j\downarrow})$ with $B_{j}=\pm B$ depending on the sublattices. The value of the local moment is extrapolated to the limit $B=0$, and a nonzero value marks the existence of the magnetism. The equal-time spin correlation function is also calculated, which is given by
\begin{equation}\label{eq4}
c_{spin}({\bf i})=\langle S^z_{{\bf j+i}}S^z_{{\bf j}}+\frac{1}{2}(S^{+}_{{\bf j+i}}S^{-}_{{\bf j}}+S^{-}_{{\bf j+i}}S^{+}_{{\bf j}})\rangle,
\end{equation}
where the spin raising and lowing operators are $S^{+}_{\bf j}=c^{\dagger}_{\bf j\uparrow}c_{\bf j\downarrow}, S^{-}_{\bf j}=c^{\dagger}_{\bf j\downarrow}c_{\bf j\uparrow}$, respectively.

\section{Magnetism of the effective model for the topologically-designed GNRs}
We first study the magnetic property based on the effective two-band Hubbard model, which writes as,
\begin{equation}\label{eq5}
H_{eff}=-\sum_{j\sigma}
(t_{j} c^\dag_{j\sigma}c^{\phantom{\dag}}_{j+1\sigma}+H.c.)+H_{U},
\end{equation}
where $t_{j}=t_1(t_2)$ for $j$ in $A(B)$ sublattice.
It is helpful to perform a MF analysis. In the MF approximation, the interaction term in Eq.(5) is decoupled as,
\begin{equation}\label{eq6}
n_{i\uparrow}n_{i\downarrow}=\langle n_{i\downarrow}\rangle n_{i\uparrow}+\langle n_{i\uparrow}\rangle n_{i\downarrow}-\langle n_{i\downarrow}\rangle \langle n_{i\uparrow} \rangle.
\end{equation}
To incorporate an antiferromagnetism order, we write
\begin{eqnarray}\label{eq7}
&&\langle n_{i\uparrow,A}\rangle=\frac{1}{2}+m_{A},\langle n_{i\downarrow,A}\rangle=\frac{1}{2}-m_{A} \\ \nonumber
&&\langle n_{i\uparrow,B}\rangle=\frac{1}{2}-m_{B},\langle n_{i\downarrow,B}\rangle=\frac{1}{2}+m_{B}.
\end{eqnarray}
Then the Hubbard term becomes,
\begin{eqnarray}\label{eq8}
&&U\sum_{i} n_{i\uparrow}n_{i\downarrow}=E_{0}+ \\ \nonumber
&&\sum_{i\in A}(-m_A n_{i\uparrow}+m_A n_{i\downarrow})+\sum_{i\in B}(m_B n_{i\uparrow}-m_B n_{i\downarrow}),
\end{eqnarray}
where $E_0=\frac{1}{4}NU+\frac{NU}{2}(m^2_A+m^2_B)$ ($N$ is the total number of the sites).
The MF Hamiltonian writes as,
\begin{align}\label{eq9}
{\cal H}^{\sigma}_{mf}({k}) =
&\left(
\begin{array}{cc}
\mp \sigma m_A & t_1+t_2 e^{-i k} \\
 t_1+t_2 e^{i k}  & \pm \sigma m_B
\end{array} \right)
\end{align}
where $\sigma=1(-1)$ represents up (down) spin.
For a uniform antiferromagnetic order, we suppose $m_{A}=m_{B}=m$. The energy spectrum is directly obtained by diagonalizing the above Hamiltonian, which is $\pm E_{k}$ with
$E_{k}=\sqrt{(U m)^2+(t_2\sin k)^2+(t_1+t_2\cos k)^2}$ (the dispersion is degenerate for both spin copies).
Minimizing the free energy $F=-2\sum_{k}E_{k}+E_0$, the self-consistent equation for the order parameter $m$ is $1=\frac{U}{N}\sum_{k}\frac{1}{E_{k}}$.

Figure 3 shows the self-consistent order parameter as a function of $U$. The value of the order parameter $m$ is zero for small interactions and the 1D chain is nonmagnetic. From a critical interaction, we have a finite self-consistent solution for $m$, implying the magnetism develops in the system. As $U$ is increased, $m$ also increases and the magnetism is strengthened. As shown in Fig.3, the curve $\frac{d m}{d U}$ tends to be divergent at the critical interaction, from which we get $U_c=0.29t$.

\begin{figure}[htbp]
\centering \includegraphics[width=7cm]{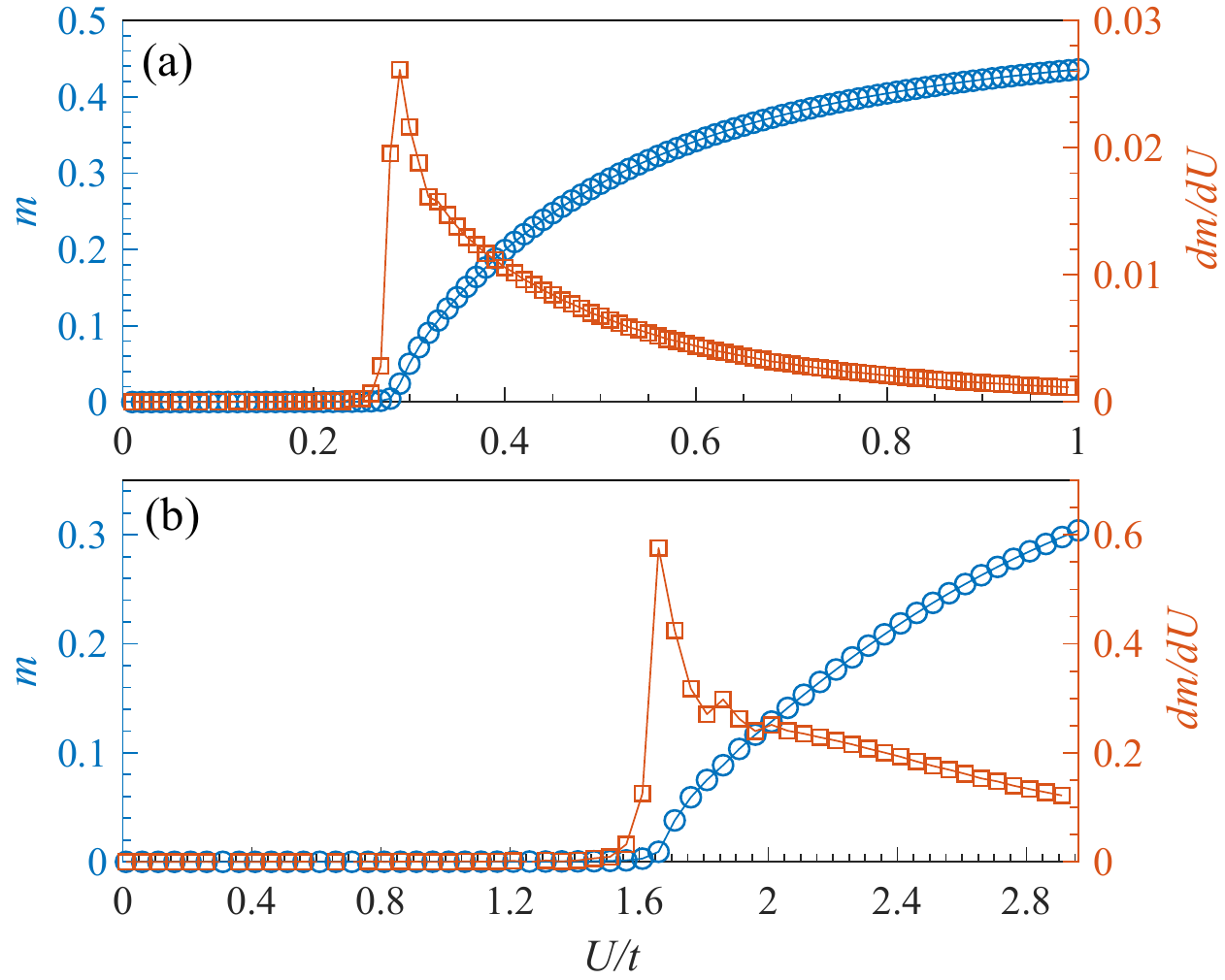} \caption{The MF order parameter $m$ as a function of $U$ based on: (a) the effective topological bands; (b) the topologically-designed GNR. Differentiating $m$ with respect to $U$, we have the critical interaction (a) $U_c=0.29t$ and (b) $U_c=1.66t$.}
\label{fig2}
\end{figure}

While the MF study provides a qualitative understanding of the magnetic behavior, the quantum fluctuation is strong in 1D, and the MF approximation may be not accurate. So we next perform exact numerical simulations on the Hubbard model in Eq.(\ref{eq5}). The equal-time
spin correlation function of the ground state is calculated using DMRG, and the result at $U=4t$ is shown in Fig.~4. The spin correlations decay with the distance, and the signs of their values are alternating, implying the correlations are antiferromagnetic. In log plot, the curve for each sublattice slightly deviates from a straight line, suggesting that the decay is slightly slower than an exponential law. It has been known that the antiferromagnetic correlation is critical for a homogeneous Hubbard chain\cite{thereza2000}. The difference may be due to the 1D chain with alternating hopping amplitudes has a gapped spectrum. The spin correlations at finite temperatures are also obtained using DQMC. In Fig.~5, we plot the spin correlation between the NN sites as a function of temperature. It shows that the curves become saturated at low temperatures, and the values steadily tend to those obtained with DMRG at zero temperature. The consistence further validates the accuracies of both methods. As the interactions are strengthened, the spin correlations also increase.
\begin{figure}[htbp]
\centering \includegraphics[width=7.cm]{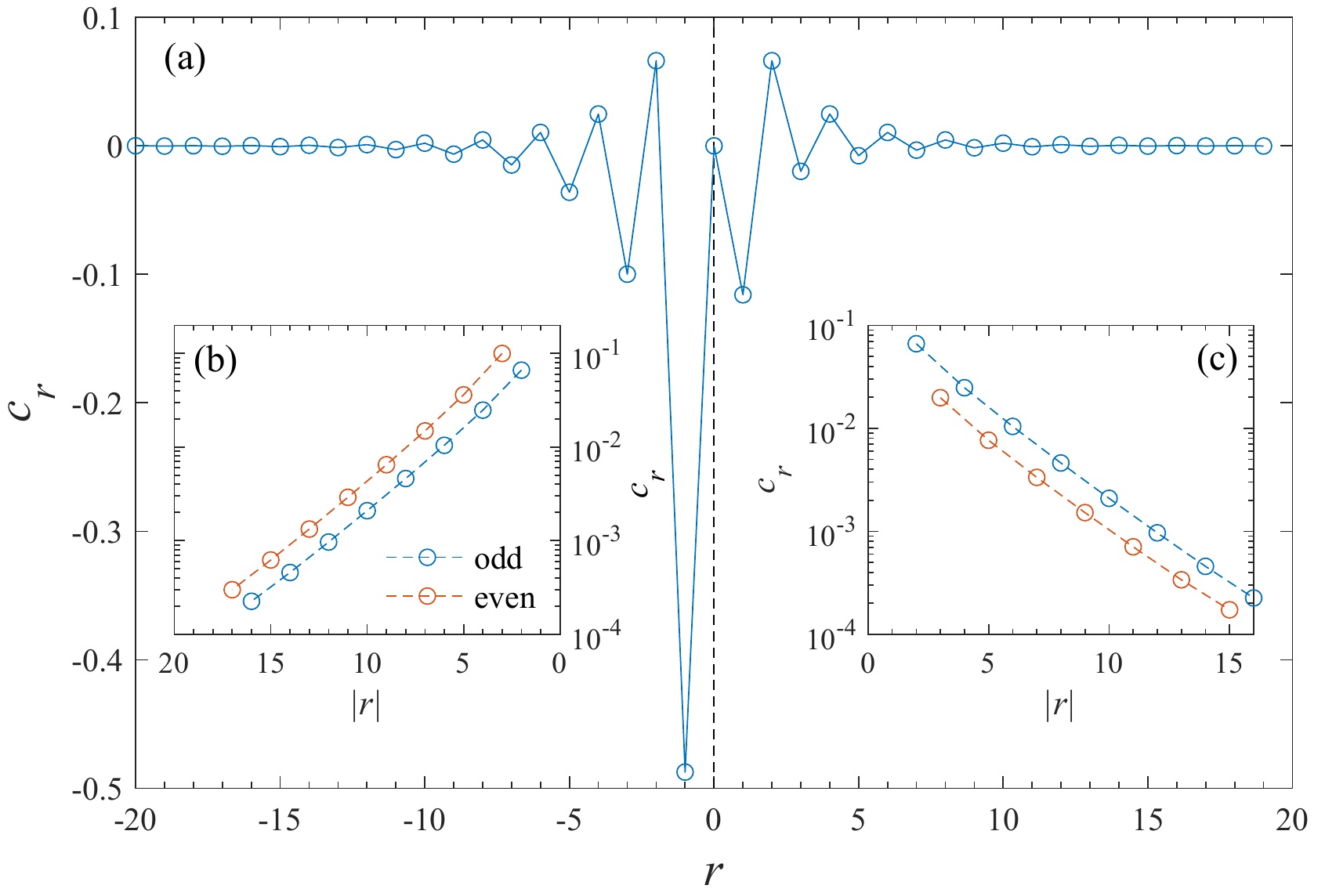} \caption{The equal-time
spin correlation as a function of distance at $U=4t$. Insets are the log plots for the left (b) and right (c) parts. The 1D lattice has $40$ sites.}
\label{fig4}
\end{figure}

\begin{figure}[htbp]
\centering \includegraphics[width=8.cm]{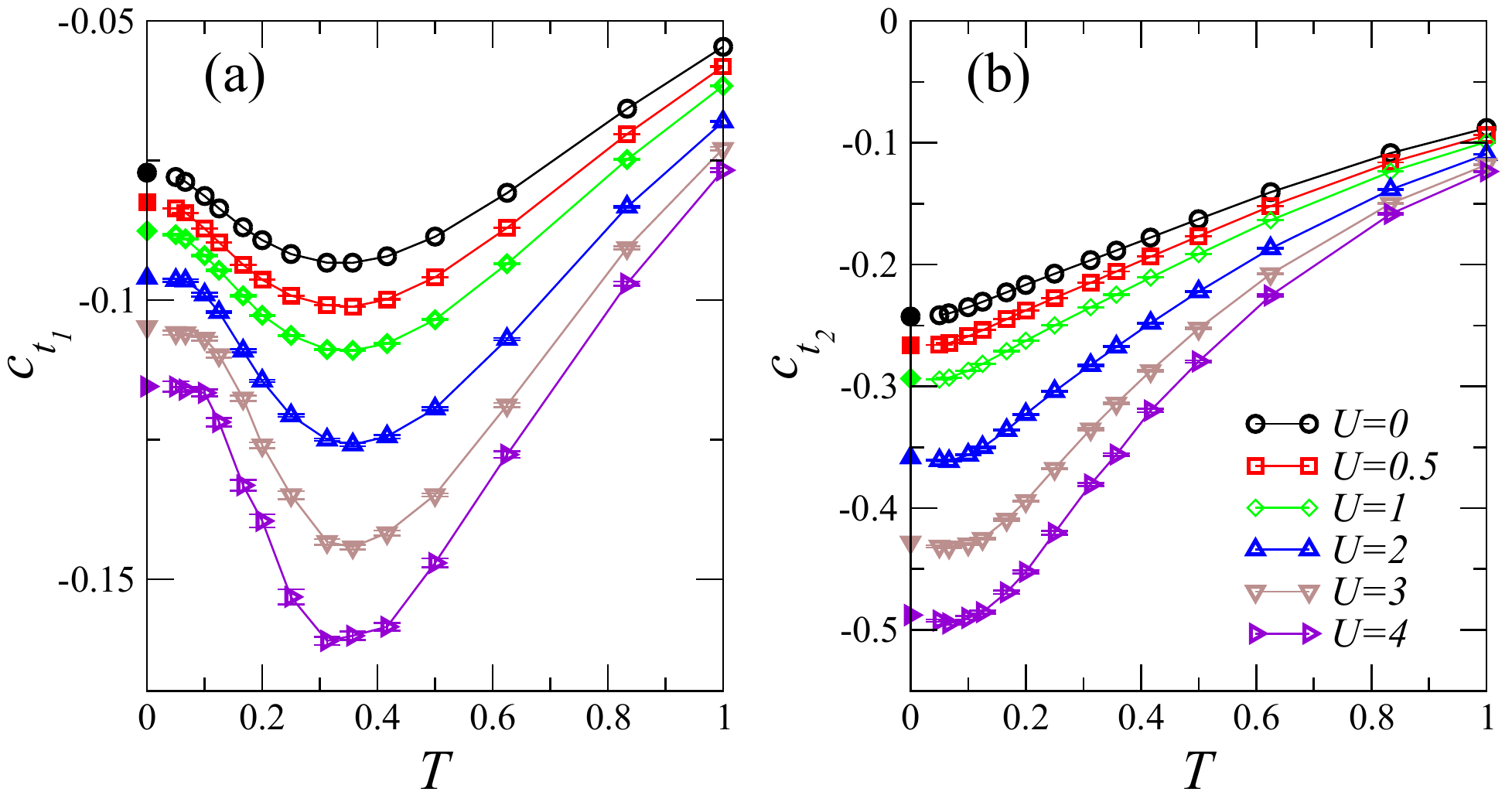} \caption{The equal-time
spin correlation function between the NN sites connected by the bonds with the hopping amplitude (a) $t_1$ and (b) $t_2$. The lattice size is the same as that of Fig.~\ref{fig4}.}
\label{fig5}
\end{figure}

\begin{figure}[htbp]
\centering \includegraphics[width=8.cm]{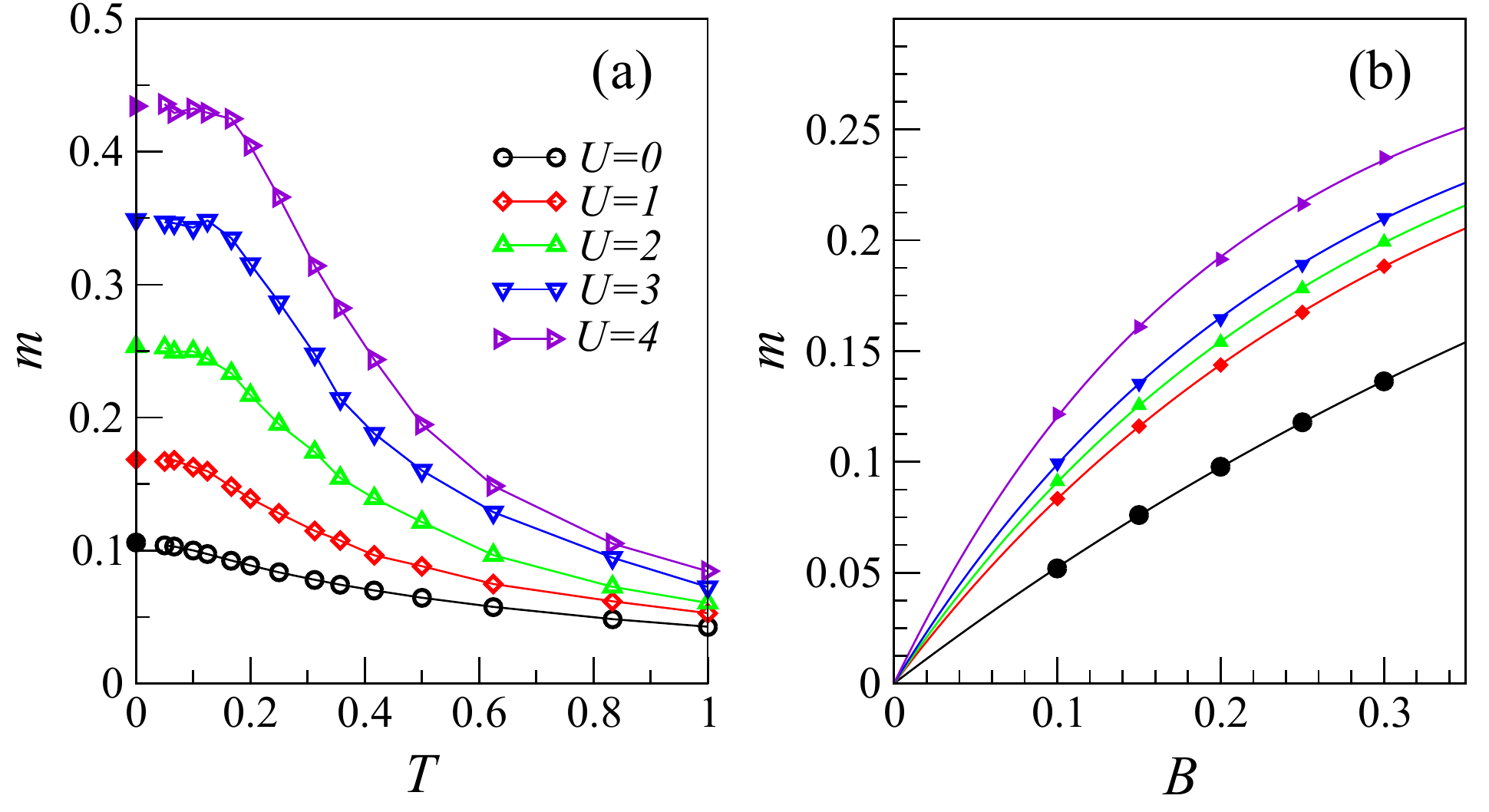} \caption{(a) The local moment $m$ as a function of temperature for several values of $U$. The pinning Zeeman field is along the $z-$axis with the strength $B=0.1$. (b) The local moment $m$ as a function of $B$. Solid curves are from the fitting function $y=a(1-e^{-bx})$. Open (solid) symbols in both figures represent DQMC (DMRG) results. The lattice size is the same as that of Fig.~\ref{fig4}.}
\label{fig6}
\end{figure}

Next we study the spontaneous-symmetry-breaking magnetism. Figure \ref{fig6} (a) shows the local moment as a function of temperature under a small symmetry-breaking term $B=0.1$. When the temperature is lowered, the thermal fluctuation is reduced and the local moment increases. Near the lowest temperatures $T=0.05$, accessed by our DQMC simulations, the value of $m$ begins to be saturated. The saturated values are in good agreement with those from the DMRG method. It is noted that the case of $U=0$ also becomes a magnet with $m\sim 0.1$, which is a natural result of the inclusion of a staggered Zeeman term. An extrapolation to $B=0$ is necessary to know the intrinsic magnetic property of the system. We carry out DMRG simulations with several different $B$ and the data are shown in Fig.~\ref{fig6}(b). They are well fitted using $y=a(1-e^{-bx})$, a function exponentially rising to maximum. So as large as $U=4t$, the extrapolated local moment is zero, and the long-ranged antiferromagnetic order is absent in a strictly 1D system.  It is in great contrast to the MF theory, where the antiferromagnetism develops above a critical interaction through a first-order phase transition.

\section{Magnetism of topologically-designed GNRs}

Next we study the experimentally relevant topological-designed GNRs, which usually contain several unit cells. We first perform a MF analysis, and decouple the Hubbard term in the same channel described in Eq.~(\ref{eq8}),
\begin{eqnarray}\label{eq10}
n_{i\uparrow}n_{i\downarrow}=\langle n_{i\downarrow}\rangle n_{i\uparrow}+\langle n_{i\uparrow}\rangle n_{i\downarrow}-\langle n_{i\downarrow}\rangle \langle n_{i\uparrow} \rangle \\ \nonumber
=-m_{i}n_{i\uparrow}+m_{i}n_{i\downarrow}+\frac{1}{4}+m^2_{i},
\end{eqnarray}
where the average density on each site writes as $n_{i\uparrow(\downarrow)}=\frac{1}{2}\pm m_{i}$.
We obtain the following MF Hamiltonian,
\begin{eqnarray}\label{eq11}
H_{mf}&=&-\sum_{\langle lj \rangle \sigma}
t c^\dag_{j\sigma}c^{\phantom{\dag}}_{l\sigma}+U\sum_{i}(-m_{i}n_{i\uparrow}+m_{i}n_{i\downarrow}) \\ \nonumber
&+&\frac{1}{4}NU+\frac{UN}{n_{s}}\sum_{i}m^2_{i},
\end{eqnarray}
where $N$ is the total number of sites, and each unit cell has $n_s=48$ sites. Minimizing the free energy $F$ with respect to $m_i$, we obtain a set of self-consistently equations $m_i=-\frac{n_s}{2UN}\frac{\partial F}{\partial m_i}$ ($i=1,...,n_{s}$). The order parameters $m_i$ are self-consistent solved, and the result is shown in Fig.~3(b). The order parameters become nonzero from a critical interaction $U_c=1.66t$, from which the magnetism develops in the ribbon.

\begin{figure}[htbp]
\centering \includegraphics[width=7.cm]{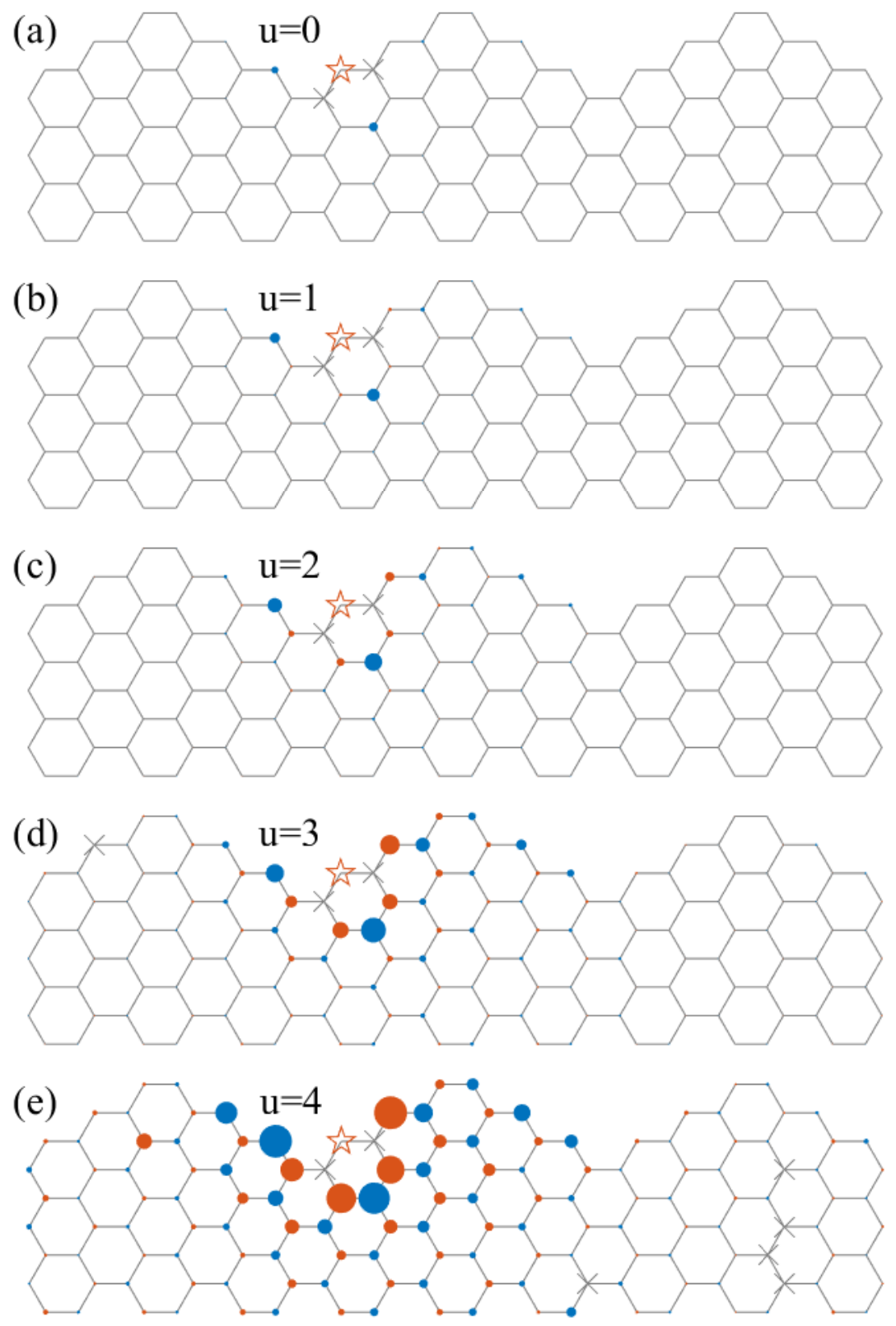} \caption{The spin correlations on a lattice containing three unit cells ($144$ sites) for several values of $U$. The star marks the reference site, and the radius of the circles denote the values of the spin correlations. The crosses mark the sites with too large values or error bars.}
\label{fig7}
\end{figure}

\begin{figure}[htbp]
\centering \includegraphics[width=7.cm]{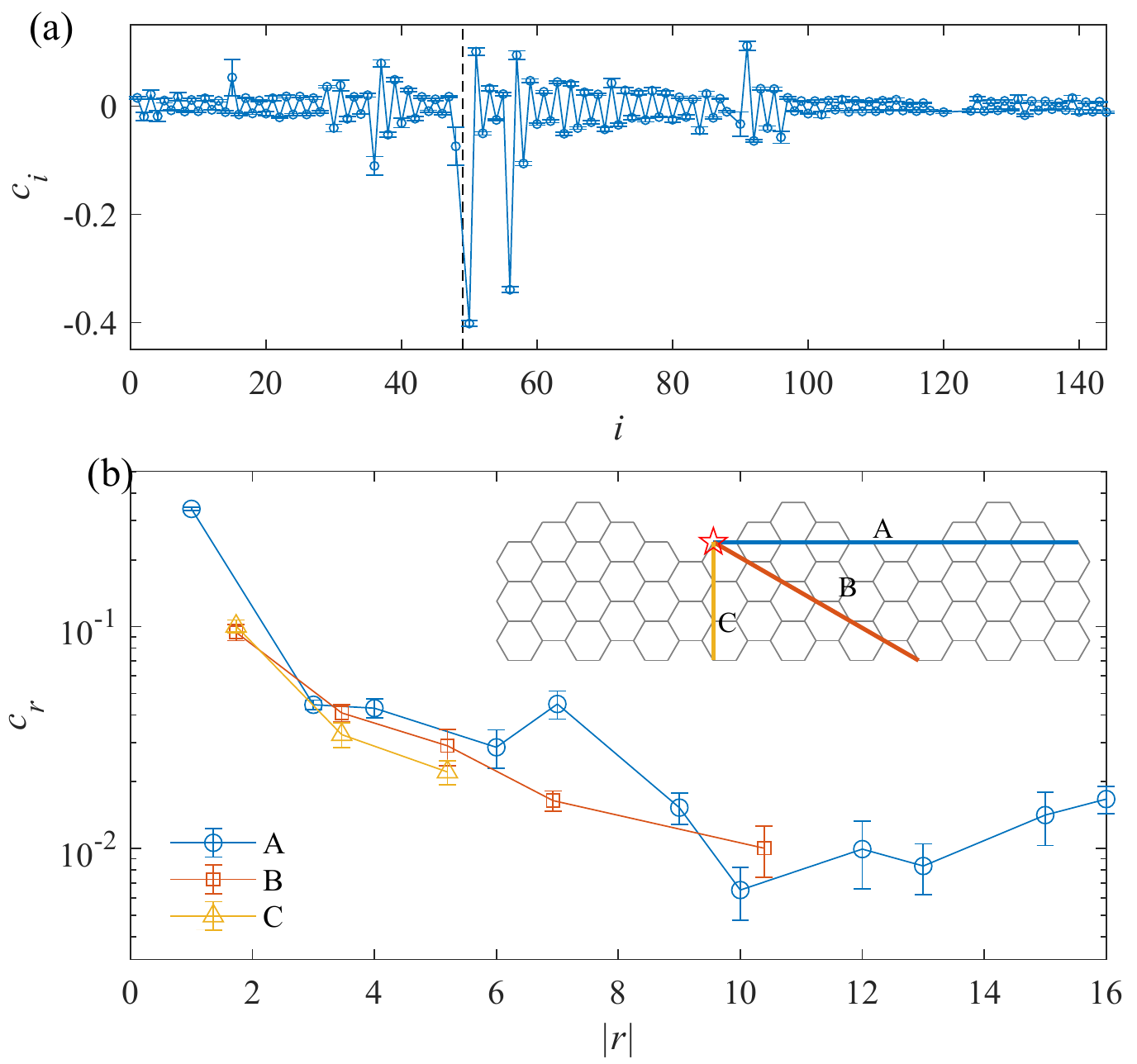} \caption{(a), The spin correlation as a function of the indexes of the sites. (b), The spin correlation as a function of the distances along several typical directions. Here the interaction is $U=4t$. The lattice size is the same as that of Fig.~\ref{fig7}.}
\label{fig8}
\end{figure}

Then the interactions are exactly dealt with using the DQMC method. Figure \ref{fig7} shows the spin correlations on a lattice containing three unit cells for several values of $U$. Their values grow as the interactions are increased, implying the interactions strengthen the spin correlations. We plot the values at $U=4$ as functions of the indexes of the sites and the distances along several typical directions. As shown in Fig.~\ref{fig8}, the spin correlations keep finite for all distances in the range of the lattice, implying the magnetism develops for large interactions.

\begin{figure}[htbp]
\centering \includegraphics[width=7.cm]{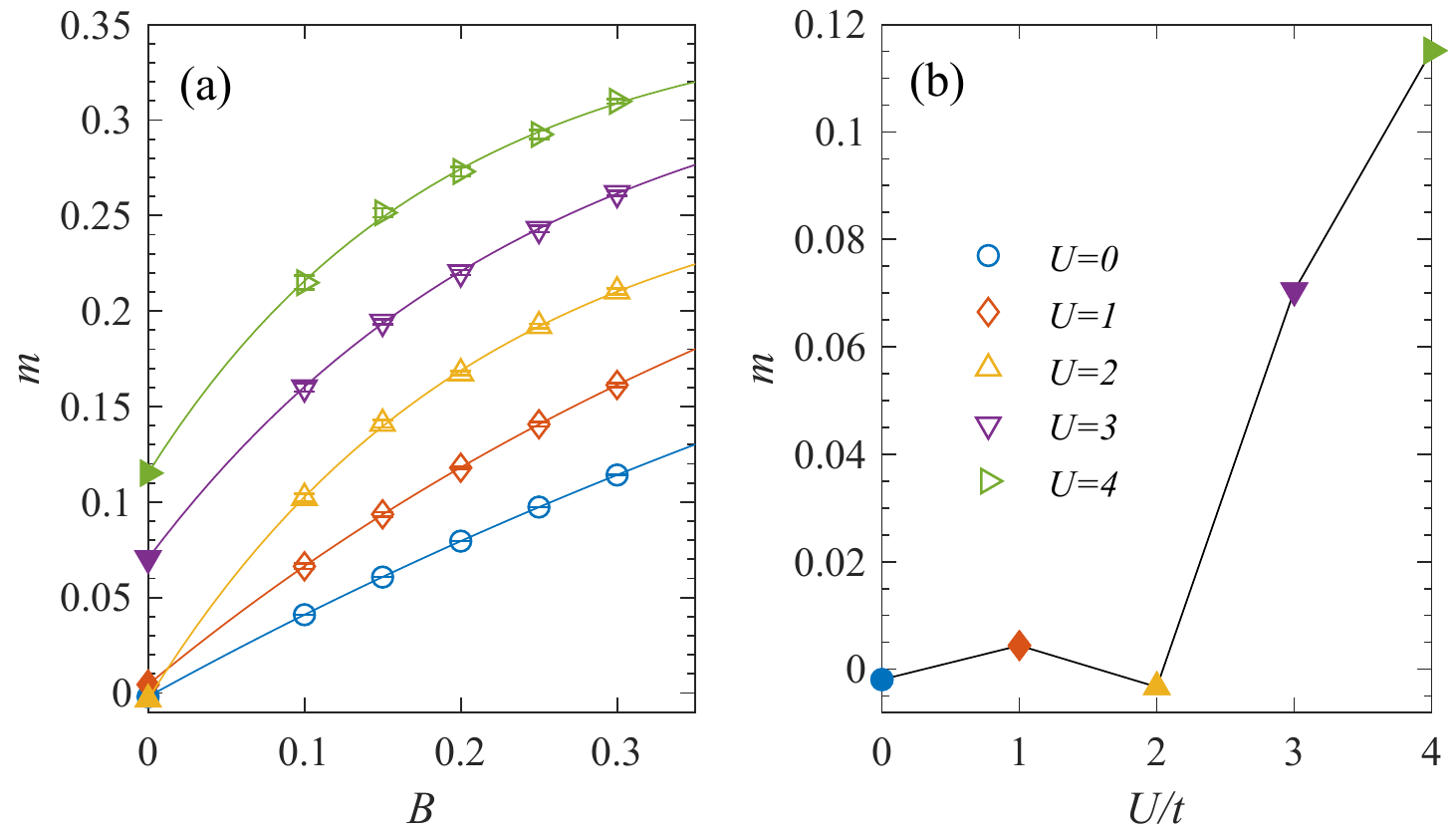} \caption{(a) The local moment $m$ as a function of $B$. Solid curves are from the fitting formula $y=c+a(1-e^{-bx})$. (b) The extrapolated local moment as a function of $U$. Here the temperature is $T=0.05t$. The lattice size is the same as that of Fig.~\ref{fig7}.}
\label{fig9}
\end{figure}

Next we study the intrinsic magnetism of the system by extrapolating a symmetry-breaking term to its vanishing limit. As shown in Fig.~\ref{fig9}(a), the local moment as a function of $B$ is best fitted using an exponentially-increasing formula $y=c+a(1-e^{-bx})$. The extrapolated local moments are almost vanishing for weak interactions, and then increase rapidly from a critical interaction. The behavior is consistent with the MF approximation. The critical interaction estimated is between $2t$ and $3t$, which is larger than the MF one. Since the realistic value of the Hubbard interaction in graphene is about $t$, the experimentally relevant GNRs should be nonmagnetic, which is consistent with the experimental results\cite{groning2018,rizzo2018}.

\section{Conclusions}
Quantum magnetism of topologically-designed GNRs is studied based on the Hubbard models using exact numerical simulations. We first study a two-band Hubbard model describing the low-energy topological bands using DMRG and DQMC methods, and show that it is nonmagnetic. We then include the Hubbard interaction to the topological-designed GNRs, and find the local moments develop from a critical interaction between $2t$ and $3t$. Compared to the realistic value in graphene, we conclude that the experimentally relevant GNRs are nonmagnetic, which is consistent with the experimental results.

\section{Acknowledgments} The work is supported by the National Key Research and Development Program of China under Grant No. 2016YFA0300304, and NSFC under Grant Nos. 11774019, 11574032 and 11734002.

\bibliography{nanoribbon_ref}

\begin{thebibliography}{39}
\expandafter\ifx\csname natexlab\endcsname\relax\def\natexlab#1{#1}\fi
\expandafter\ifx\csname bibnamefont\endcsname\relax
  \def\bibnamefont#1{#1}\fi
\expandafter\ifx\csname bibfnamefont\endcsname\relax
  \def\bibfnamefont#1{#1}\fi
\expandafter\ifx\csname citenamefont\endcsname\relax
  \def\citenamefont#1{#1}\fi
\expandafter\ifx\csname url\endcsname\relax
  \def\url#1{\texttt{#1}}\fi
\expandafter\ifx\csname urlprefix\endcsname\relax\def\urlprefix{URL }\fi
\providecommand{\bibinfo}[2]{#2}
\providecommand{\eprint}[2][]{\url{#2}}

\bibitem[{\citenamefont{Yazyev}(2010)}]{yazyev2010}
\bibinfo{author}{\bibfnamefont{O.~V.} \bibnamefont{Yazyev}},
  \bibinfo{journal}{Reports on Progress in Physics}
  \textbf{\bibinfo{volume}{73}}, \bibinfo{pages}{056501}
  (\bibinfo{year}{2010}).

\bibitem[{\citenamefont{Son et~al.}(2006{\natexlab{a}})\citenamefont{Son,
  Cohen, and Louie}}]{son2006b}
\bibinfo{author}{\bibfnamefont{Y.}~\bibnamefont{Son}},
  \bibinfo{author}{\bibfnamefont{M.~L.} \bibnamefont{Cohen}}, \bibnamefont{and}
  \bibinfo{author}{\bibfnamefont{S.~G.} \bibnamefont{Louie}},
  \bibinfo{journal}{Nature} \textbf{\bibinfo{volume}{444}},
  \bibinfo{pages}{347} (\bibinfo{year}{2006}{\natexlab{a}}).

\bibitem[{\citenamefont{Yazyev}(2013)}]{yazyev2013}
\bibinfo{author}{\bibfnamefont{O.~V.} \bibnamefont{Yazyev}},
  \bibinfo{journal}{Accounts of Chemical Research}
  \textbf{\bibinfo{volume}{46}}, \bibinfo{pages}{2319} (\bibinfo{year}{2013}).

\bibitem[{\citenamefont{Nakada et~al.}(1996)\citenamefont{Nakada, Fujita,
  Dresselhaus, and Dresselhaus}}]{nakada1996}
\bibinfo{author}{\bibfnamefont{K.}~\bibnamefont{Nakada}},
  \bibinfo{author}{\bibfnamefont{M.}~\bibnamefont{Fujita}},
  \bibinfo{author}{\bibfnamefont{G.}~\bibnamefont{Dresselhaus}},
  \bibnamefont{and} \bibinfo{author}{\bibfnamefont{M.~S.}
  \bibnamefont{Dresselhaus}}, \bibinfo{journal}{Phys. Rev. B}
  \textbf{\bibinfo{volume}{54}}, \bibinfo{pages}{17954} (\bibinfo{year}{1996}),
  \urlprefix\url{https://link.aps.org/doi/10.1103/PhysRevB.54.17954}.

\bibitem[{\citenamefont{Kimouche et~al.}(2015)\citenamefont{Kimouche, Ervasti,
  Drost, Halonen, Harju, Joensuu, Sainio, and Liljeroth}}]{kimouche2015}
\bibinfo{author}{\bibfnamefont{A.}~\bibnamefont{Kimouche}},
  \bibinfo{author}{\bibfnamefont{M.~M.} \bibnamefont{Ervasti}},
  \bibinfo{author}{\bibfnamefont{R.~J.} \bibnamefont{Drost}},
  \bibinfo{author}{\bibfnamefont{S.}~\bibnamefont{Halonen}},
  \bibinfo{author}{\bibfnamefont{A.}~\bibnamefont{Harju}},
  \bibinfo{author}{\bibfnamefont{P.}~\bibnamefont{Joensuu}},
  \bibinfo{author}{\bibfnamefont{J.}~\bibnamefont{Sainio}}, \bibnamefont{and}
  \bibinfo{author}{\bibfnamefont{P.}~\bibnamefont{Liljeroth}},
  \bibinfo{journal}{Nature Communications} \textbf{\bibinfo{volume}{6}},
  \bibinfo{pages}{10177} (\bibinfo{year}{2015}).

\bibitem[{\citenamefont{Ruffieux et~al.}(2016)\citenamefont{Ruffieux, Wang,
  Yang, Sanchezsanchez, Liu, Dienel, Talirz, Shinde, Pignedoli, Passerone
  et~al.}}]{ruffieux2016}
\bibinfo{author}{\bibfnamefont{P.}~\bibnamefont{Ruffieux}},
  \bibinfo{author}{\bibfnamefont{S.}~\bibnamefont{Wang}},
  \bibinfo{author}{\bibfnamefont{B.}~\bibnamefont{Yang}},
  \bibinfo{author}{\bibfnamefont{C.~M.} \bibnamefont{Sanchezsanchez}},
  \bibinfo{author}{\bibfnamefont{J.}~\bibnamefont{Liu}},
  \bibinfo{author}{\bibfnamefont{T.}~\bibnamefont{Dienel}},
  \bibinfo{author}{\bibfnamefont{L.}~\bibnamefont{Talirz}},
  \bibinfo{author}{\bibfnamefont{P.}~\bibnamefont{Shinde}},
  \bibinfo{author}{\bibfnamefont{C.~A.} \bibnamefont{Pignedoli}},
  \bibinfo{author}{\bibfnamefont{D.}~\bibnamefont{Passerone}},
  \bibnamefont{et~al.}, \bibinfo{journal}{Nature}
  \textbf{\bibinfo{volume}{531}}, \bibinfo{pages}{489} (\bibinfo{year}{2016}).

\bibitem[{\citenamefont{Castro~Neto et~al.}(2009)\citenamefont{Castro~Neto,
  Guinea, Peres, Novoselov, and Geim}}]{castro2009}
\bibinfo{author}{\bibfnamefont{A.~H.} \bibnamefont{Castro~Neto}},
  \bibinfo{author}{\bibfnamefont{F.}~\bibnamefont{Guinea}},
  \bibinfo{author}{\bibfnamefont{N.~M.~R.} \bibnamefont{Peres}},
  \bibinfo{author}{\bibfnamefont{K.~S.} \bibnamefont{Novoselov}},
  \bibnamefont{and} \bibinfo{author}{\bibfnamefont{A.~K.} \bibnamefont{Geim}},
  \bibinfo{journal}{Rev. Mod. Phys.} \textbf{\bibinfo{volume}{81}},
  \bibinfo{pages}{109} (\bibinfo{year}{2009}),
  \urlprefix\url{https://link.aps.org/doi/10.1103/RevModPhys.81.109}.

\bibitem[{\citenamefont{Sorella and Tosatti}(1992)}]{sorella1992}
\bibinfo{author}{\bibfnamefont{S.}~\bibnamefont{Sorella}} \bibnamefont{and}
  \bibinfo{author}{\bibfnamefont{E.}~\bibnamefont{Tosatti}},
  \bibinfo{journal}{Europhysics Letters ({EPL})} \textbf{\bibinfo{volume}{19}},
  \bibinfo{pages}{699} (\bibinfo{year}{1992}),
  \urlprefix\url{https://doi.org/10.1209%2F0295-5075%2F19%2F8%2F007}.

\bibitem[{\citenamefont{Paiva et~al.}(2005)\citenamefont{Paiva, Scalettar,
  Zheng, Singh, and Oitmaa}}]{paiva2005}
\bibinfo{author}{\bibfnamefont{T.}~\bibnamefont{Paiva}},
  \bibinfo{author}{\bibfnamefont{R.~T.} \bibnamefont{Scalettar}},
  \bibinfo{author}{\bibfnamefont{W.}~\bibnamefont{Zheng}},
  \bibinfo{author}{\bibfnamefont{R.~R.~P.} \bibnamefont{Singh}},
  \bibnamefont{and} \bibinfo{author}{\bibfnamefont{J.}~\bibnamefont{Oitmaa}},
  \bibinfo{journal}{Phys. Rev. B} \textbf{\bibinfo{volume}{72}},
  \bibinfo{pages}{085123} (\bibinfo{year}{2005}),
  \urlprefix\url{https://link.aps.org/doi/10.1103/PhysRevB.72.085123}.

\bibitem[{\citenamefont{Meng et~al.}(2010)\citenamefont{Meng, Lang, Wessel,
  Assaad, and Muramatsu}}]{meng2010}
\bibinfo{author}{\bibfnamefont{Z.}~\bibnamefont{Meng}},
  \bibinfo{author}{\bibfnamefont{T.}~\bibnamefont{Lang}},
  \bibinfo{author}{\bibfnamefont{S.}~\bibnamefont{Wessel}},
  \bibinfo{author}{\bibfnamefont{F.}~\bibnamefont{Assaad}}, \bibnamefont{and}
  \bibinfo{author}{\bibfnamefont{A.}~\bibnamefont{Muramatsu}},
  \bibinfo{journal}{Nature} \textbf{\bibinfo{volume}{464}},
  \bibinfo{pages}{847} (\bibinfo{year}{2010}).

\bibitem[{\citenamefont{Sorella et~al.}(2012)\citenamefont{Sorella, Otsuka, and
  Yunoki}}]{sorella2012}
\bibinfo{author}{\bibfnamefont{S.}~\bibnamefont{Sorella}},
  \bibinfo{author}{\bibfnamefont{Y.}~\bibnamefont{Otsuka}}, \bibnamefont{and}
  \bibinfo{author}{\bibfnamefont{S.}~\bibnamefont{Yunoki}},
  \bibinfo{journal}{Scientific reports} \textbf{\bibinfo{volume}{2}},
  \bibinfo{pages}{992} (\bibinfo{year}{2012}).

\bibitem[{\citenamefont{Feldner et~al.}(2011)\citenamefont{Feldner, Meng, Lang,
  Assaad, Wessel, and Honecker}}]{feldner2011}
\bibinfo{author}{\bibfnamefont{H.}~\bibnamefont{Feldner}},
  \bibinfo{author}{\bibfnamefont{Z.~Y.} \bibnamefont{Meng}},
  \bibinfo{author}{\bibfnamefont{T.~C.} \bibnamefont{Lang}},
  \bibinfo{author}{\bibfnamefont{F.~F.} \bibnamefont{Assaad}},
  \bibinfo{author}{\bibfnamefont{S.}~\bibnamefont{Wessel}}, \bibnamefont{and}
  \bibinfo{author}{\bibfnamefont{A.}~\bibnamefont{Honecker}},
  \bibinfo{journal}{Phys. Rev. Lett.} \textbf{\bibinfo{volume}{106}},
  \bibinfo{pages}{226401} (\bibinfo{year}{2011}),
  \urlprefix\url{https://link.aps.org/doi/10.1103/PhysRevLett.106.226401}.

\bibitem[{\citenamefont{Hikihara et~al.}(2003)\citenamefont{Hikihara, Hu, Lin,
  and Mou}}]{hikihara2003}
\bibinfo{author}{\bibfnamefont{T.}~\bibnamefont{Hikihara}},
  \bibinfo{author}{\bibfnamefont{X.}~\bibnamefont{Hu}},
  \bibinfo{author}{\bibfnamefont{H.-H.} \bibnamefont{Lin}}, \bibnamefont{and}
  \bibinfo{author}{\bibfnamefont{C.-Y.} \bibnamefont{Mou}},
  \bibinfo{journal}{Phys. Rev. B} \textbf{\bibinfo{volume}{68}},
  \bibinfo{pages}{035432} (\bibinfo{year}{2003}),
  \urlprefix\url{https://link.aps.org/doi/10.1103/PhysRevB.68.035432}.

\bibitem[{\citenamefont{Feldner et~al.}(2010)\citenamefont{Feldner, Meng,
  Honecker, Cabra, Wessel, and Assaad}}]{feldner2010}
\bibinfo{author}{\bibfnamefont{H.}~\bibnamefont{Feldner}},
  \bibinfo{author}{\bibfnamefont{Z.~Y.} \bibnamefont{Meng}},
  \bibinfo{author}{\bibfnamefont{A.}~\bibnamefont{Honecker}},
  \bibinfo{author}{\bibfnamefont{D.}~\bibnamefont{Cabra}},
  \bibinfo{author}{\bibfnamefont{S.}~\bibnamefont{Wessel}}, \bibnamefont{and}
  \bibinfo{author}{\bibfnamefont{F.~F.} \bibnamefont{Assaad}},
  \bibinfo{journal}{Phys. Rev. B} \textbf{\bibinfo{volume}{81}},
  \bibinfo{pages}{115416} (\bibinfo{year}{2010}),
  \urlprefix\url{https://link.aps.org/doi/10.1103/PhysRevB.81.115416}.

\bibitem[{\citenamefont{Luitz et~al.}(2011)\citenamefont{Luitz, Assaad, and
  Schmidt}}]{luitz2011}
\bibinfo{author}{\bibfnamefont{D.~J.} \bibnamefont{Luitz}},
  \bibinfo{author}{\bibfnamefont{F.~F.} \bibnamefont{Assaad}},
  \bibnamefont{and} \bibinfo{author}{\bibfnamefont{M.~J.}
  \bibnamefont{Schmidt}}, \bibinfo{journal}{Phys. Rev. B}
  \textbf{\bibinfo{volume}{83}}, \bibinfo{pages}{195432}
  (\bibinfo{year}{2011}),
  \urlprefix\url{https://link.aps.org/doi/10.1103/PhysRevB.83.195432}.

\bibitem[{\citenamefont{Schmidt and Loss}(2010)}]{schmidt2010}
\bibinfo{author}{\bibfnamefont{M.~J.} \bibnamefont{Schmidt}} \bibnamefont{and}
  \bibinfo{author}{\bibfnamefont{D.}~\bibnamefont{Loss}},
  \bibinfo{journal}{Phys. Rev. B} \textbf{\bibinfo{volume}{82}},
  \bibinfo{pages}{085422} (\bibinfo{year}{2010}),
  \urlprefix\url{https://link.aps.org/doi/10.1103/PhysRevB.82.085422}.

\bibitem[{\citenamefont{Golor et~al.}(2013)\citenamefont{Golor, Lang, and
  Wessel}}]{wessel2013}
\bibinfo{author}{\bibfnamefont{M.}~\bibnamefont{Golor}},
  \bibinfo{author}{\bibfnamefont{T.~C.} \bibnamefont{Lang}}, \bibnamefont{and}
  \bibinfo{author}{\bibfnamefont{S.}~\bibnamefont{Wessel}},
  \bibinfo{journal}{Phys. Rev. B} \textbf{\bibinfo{volume}{87}},
  \bibinfo{pages}{155441} (\bibinfo{year}{2013}),
  \urlprefix\url{https://link.aps.org/doi/10.1103/PhysRevB.87.155441}.

\bibitem[{\citenamefont{Hagym\'asi and Legeza}(2016)}]{hagrmasi2016}
\bibinfo{author}{\bibfnamefont{I.}~\bibnamefont{Hagym\'asi}} \bibnamefont{and}
  \bibinfo{author}{\bibfnamefont{O.}~\bibnamefont{Legeza}},
  \bibinfo{journal}{Phys. Rev. B} \textbf{\bibinfo{volume}{94}},
  \bibinfo{pages}{165147} (\bibinfo{year}{2016}),
  \urlprefix\url{https://link.aps.org/doi/10.1103/PhysRevB.94.165147}.

\bibitem[{\citenamefont{Joly et~al.}(2010)\citenamefont{Joly, Kiguchi, Hao,
  Takai, Enoki, Sumii, Amemiya, Muramatsu, Hayashi, Kim et~al.}}]{joly2010}
\bibinfo{author}{\bibfnamefont{V.~L.~J.} \bibnamefont{Joly}},
  \bibinfo{author}{\bibfnamefont{M.}~\bibnamefont{Kiguchi}},
  \bibinfo{author}{\bibfnamefont{S.-J.} \bibnamefont{Hao}},
  \bibinfo{author}{\bibfnamefont{K.}~\bibnamefont{Takai}},
  \bibinfo{author}{\bibfnamefont{T.}~\bibnamefont{Enoki}},
  \bibinfo{author}{\bibfnamefont{R.}~\bibnamefont{Sumii}},
  \bibinfo{author}{\bibfnamefont{K.}~\bibnamefont{Amemiya}},
  \bibinfo{author}{\bibfnamefont{H.}~\bibnamefont{Muramatsu}},
  \bibinfo{author}{\bibfnamefont{T.}~\bibnamefont{Hayashi}},
  \bibinfo{author}{\bibfnamefont{Y.~A.} \bibnamefont{Kim}},
  \bibnamefont{et~al.}, \bibinfo{journal}{Phys. Rev. B}
  \textbf{\bibinfo{volume}{81}}, \bibinfo{pages}{245428}
  (\bibinfo{year}{2010}),
  \urlprefix\url{https://link.aps.org/doi/10.1103/PhysRevB.81.245428}.

\bibitem[{\citenamefont{Magda et~al.}(2014)\citenamefont{Magda, Jin, Hagymasi,
  Vancso, Osvath, Nemesincze, Hwang, Biro, and Tapaszto}}]{magda2014}
\bibinfo{author}{\bibfnamefont{G.~Z.} \bibnamefont{Magda}},
  \bibinfo{author}{\bibfnamefont{X.}~\bibnamefont{Jin}},
  \bibinfo{author}{\bibfnamefont{I.}~\bibnamefont{Hagymasi}},
  \bibinfo{author}{\bibfnamefont{P.}~\bibnamefont{Vancso}},
  \bibinfo{author}{\bibfnamefont{Z.}~\bibnamefont{Osvath}},
  \bibinfo{author}{\bibfnamefont{P.}~\bibnamefont{Nemesincze}},
  \bibinfo{author}{\bibfnamefont{C.}~\bibnamefont{Hwang}},
  \bibinfo{author}{\bibfnamefont{L.~P.} \bibnamefont{Biro}}, \bibnamefont{and}
  \bibinfo{author}{\bibfnamefont{L.}~\bibnamefont{Tapaszto}},
  \bibinfo{journal}{Nature} \textbf{\bibinfo{volume}{514}},
  \bibinfo{pages}{608} (\bibinfo{year}{2014}).

\bibitem[{\citenamefont{Son et~al.}(2006{\natexlab{b}})\citenamefont{Son,
  Cohen, and Louie}}]{son2006}
\bibinfo{author}{\bibfnamefont{Y.-W.} \bibnamefont{Son}},
  \bibinfo{author}{\bibfnamefont{M.~L.} \bibnamefont{Cohen}}, \bibnamefont{and}
  \bibinfo{author}{\bibfnamefont{S.~G.} \bibnamefont{Louie}},
  \bibinfo{journal}{Phys. Rev. Lett.} \textbf{\bibinfo{volume}{97}},
  \bibinfo{pages}{216803} (\bibinfo{year}{2006}{\natexlab{b}}),
  \urlprefix\url{https://link.aps.org/doi/10.1103/PhysRevLett.97.216803}.

\bibitem[{\citenamefont{Brey and Fertig}(2006)}]{brey2006}
\bibinfo{author}{\bibfnamefont{L.}~\bibnamefont{Brey}} \bibnamefont{and}
  \bibinfo{author}{\bibfnamefont{H.~A.} \bibnamefont{Fertig}},
  \bibinfo{journal}{Phys. Rev. B} \textbf{\bibinfo{volume}{73}},
  \bibinfo{pages}{235411} (\bibinfo{year}{2006}),
  \urlprefix\url{https://link.aps.org/doi/10.1103/PhysRevB.73.235411}.

\bibitem[{\citenamefont{Ezawa}(2006)}]{ezawa2006}
\bibinfo{author}{\bibfnamefont{M.}~\bibnamefont{Ezawa}},
  \bibinfo{journal}{Phys. Rev. B} \textbf{\bibinfo{volume}{73}},
  \bibinfo{pages}{045432} (\bibinfo{year}{2006}),
  \urlprefix\url{https://link.aps.org/doi/10.1103/PhysRevB.73.045432}.

\bibitem[{\citenamefont{Cao et~al.}(2017)\citenamefont{Cao, Zhao, and
  Louie}}]{cao2017}
\bibinfo{author}{\bibfnamefont{T.}~\bibnamefont{Cao}},
  \bibinfo{author}{\bibfnamefont{F.}~\bibnamefont{Zhao}}, \bibnamefont{and}
  \bibinfo{author}{\bibfnamefont{S.~G.} \bibnamefont{Louie}},
  \bibinfo{journal}{Phys. Rev. Lett.} \textbf{\bibinfo{volume}{119}},
  \bibinfo{pages}{076401} (\bibinfo{year}{2017}),
  \urlprefix\url{https://link.aps.org/doi/10.1103/PhysRevLett.119.076401}.

\bibitem[{\citenamefont{Lin and Chou}(2018)}]{lin2018}
\bibinfo{author}{\bibfnamefont{K.-S.} \bibnamefont{Lin}} \bibnamefont{and}
  \bibinfo{author}{\bibfnamefont{M.-Y.} \bibnamefont{Chou}},
  \bibinfo{journal}{Nano LettersNano Letters} \textbf{\bibinfo{volume}{18}},
  \bibinfo{pages}{7254 } (\bibinfo{year}{2018}), \bibinfo{note}{doi:
  10.1021/acs.nanolett.8b03417},
  \urlprefix\url{https://doi.org/10.1021/acs.nanolett.8b03417}.

\bibitem[{\citenamefont{Groning et~al.}(2018)\citenamefont{Groning, Wang, Yao,
  Pignedoli, Barin, Daniels, Cupo, Meunier, Feng, Narita et~al.}}]{groning2018}
\bibinfo{author}{\bibfnamefont{O.}~\bibnamefont{Groning}},
  \bibinfo{author}{\bibfnamefont{S.}~\bibnamefont{Wang}},
  \bibinfo{author}{\bibfnamefont{X.}~\bibnamefont{Yao}},
  \bibinfo{author}{\bibfnamefont{C.~A.} \bibnamefont{Pignedoli}},
  \bibinfo{author}{\bibfnamefont{G.~B.} \bibnamefont{Barin}},
  \bibinfo{author}{\bibfnamefont{C.}~\bibnamefont{Daniels}},
  \bibinfo{author}{\bibfnamefont{A.}~\bibnamefont{Cupo}},
  \bibinfo{author}{\bibfnamefont{V.}~\bibnamefont{Meunier}},
  \bibinfo{author}{\bibfnamefont{X.}~\bibnamefont{Feng}},
  \bibinfo{author}{\bibfnamefont{A.}~\bibnamefont{Narita}},
  \bibnamefont{et~al.}, \bibinfo{journal}{Nature}
  \textbf{\bibinfo{volume}{560}}, \bibinfo{pages}{209} (\bibinfo{year}{2018}).

\bibitem[{\citenamefont{Rizzo et~al.}(2018)\citenamefont{Rizzo, Veber, Cao,
  Bronner, Chen, Zhao, Rodriguez, Louie, Crommie, and Fischer}}]{rizzo2018}
\bibinfo{author}{\bibfnamefont{D.~J.} \bibnamefont{Rizzo}},
  \bibinfo{author}{\bibfnamefont{G.}~\bibnamefont{Veber}},
  \bibinfo{author}{\bibfnamefont{T.}~\bibnamefont{Cao}},
  \bibinfo{author}{\bibfnamefont{C.}~\bibnamefont{Bronner}},
  \bibinfo{author}{\bibfnamefont{T.}~\bibnamefont{Chen}},
  \bibinfo{author}{\bibfnamefont{F.}~\bibnamefont{Zhao}},
  \bibinfo{author}{\bibfnamefont{H.}~\bibnamefont{Rodriguez}},
  \bibinfo{author}{\bibfnamefont{S.~G.} \bibnamefont{Louie}},
  \bibinfo{author}{\bibfnamefont{M.~F.} \bibnamefont{Crommie}},
  \bibnamefont{and} \bibinfo{author}{\bibfnamefont{F.~R.}
  \bibnamefont{Fischer}}, \bibinfo{journal}{Nature}
  \textbf{\bibinfo{volume}{560}}, \bibinfo{pages}{204} (\bibinfo{year}{2018}).

\bibitem[{\citenamefont{Su et~al.}(1979)\citenamefont{Su, Schrieffer, and
  Heeger}}]{su1979}
\bibinfo{author}{\bibfnamefont{W.~P.} \bibnamefont{Su}},
  \bibinfo{author}{\bibfnamefont{J.~R.} \bibnamefont{Schrieffer}},
  \bibnamefont{and} \bibinfo{author}{\bibfnamefont{A.~J.}
  \bibnamefont{Heeger}}, \bibinfo{journal}{Phys. Rev. Lett.}
  \textbf{\bibinfo{volume}{42}}, \bibinfo{pages}{1698} (\bibinfo{year}{1979}),
  \urlprefix\url{https://link.aps.org/doi/10.1103/PhysRevLett.42.1698}.

\bibitem[{\citenamefont{Marzari et~al.}(2012)\citenamefont{Marzari, Mostofi,
  Yates, Souza, and Vanderbilt}}]{marzari2012}
\bibinfo{author}{\bibfnamefont{N.}~\bibnamefont{Marzari}},
  \bibinfo{author}{\bibfnamefont{A.~A.} \bibnamefont{Mostofi}},
  \bibinfo{author}{\bibfnamefont{J.~R.} \bibnamefont{Yates}},
  \bibinfo{author}{\bibfnamefont{I.}~\bibnamefont{Souza}}, \bibnamefont{and}
  \bibinfo{author}{\bibfnamefont{D.}~\bibnamefont{Vanderbilt}},
  \bibinfo{journal}{Rev. Mod. Phys.} \textbf{\bibinfo{volume}{84}},
  \bibinfo{pages}{1419} (\bibinfo{year}{2012}),
  \urlprefix\url{https://link.aps.org/doi/10.1103/RevModPhys.84.1419}.

\bibitem[{\citenamefont{Guo and Shen}(2011)}]{berry1}
\bibinfo{author}{\bibfnamefont{H.}~\bibnamefont{Guo}} \bibnamefont{and}
  \bibinfo{author}{\bibfnamefont{S.-Q.} \bibnamefont{Shen}},
  \bibinfo{journal}{Phys. Rev. B} \textbf{\bibinfo{volume}{84}},
  \bibinfo{pages}{195107} (\bibinfo{year}{2011}),
  \urlprefix\url{https://link.aps.org/doi/10.1103/PhysRevB.84.195107}.

\bibitem[{\citenamefont{Guo et~al.}(2012)\citenamefont{Guo, Shen, and
  Feng}}]{berry2}
\bibinfo{author}{\bibfnamefont{H.}~\bibnamefont{Guo}},
  \bibinfo{author}{\bibfnamefont{S.-Q.} \bibnamefont{Shen}}, \bibnamefont{and}
  \bibinfo{author}{\bibfnamefont{S.}~\bibnamefont{Feng}},
  \bibinfo{journal}{Phys. Rev. B} \textbf{\bibinfo{volume}{86}},
  \bibinfo{pages}{085124} (\bibinfo{year}{2012}),
  \urlprefix\url{https://link.aps.org/doi/10.1103/PhysRevB.86.085124}.

\bibitem[{\citenamefont{Wang and Zhang}(2014)}]{wang2014}
\bibinfo{author}{\bibfnamefont{Z.}~\bibnamefont{Wang}} \bibnamefont{and}
  \bibinfo{author}{\bibfnamefont{S.-C.} \bibnamefont{Zhang}},
  \bibinfo{journal}{Phys. Rev. X} \textbf{\bibinfo{volume}{4}},
  \bibinfo{pages}{011006} (\bibinfo{year}{2014}),
  \urlprefix\url{https://link.aps.org/doi/10.1103/PhysRevX.4.011006}.

\bibitem[{\citenamefont{Resta}(1994)}]{resta1994}
\bibinfo{author}{\bibfnamefont{R.}~\bibnamefont{Resta}}, \bibinfo{journal}{Rev.
  Mod. Phys.} \textbf{\bibinfo{volume}{66}}, \bibinfo{pages}{899}
  (\bibinfo{year}{1994}),
  \urlprefix\url{https://link.aps.org/doi/10.1103/RevModPhys.66.899}.

\bibitem[{\citenamefont{Xiao et~al.}(2010)\citenamefont{Xiao, Chang, and
  Niu}}]{xiao2010}
\bibinfo{author}{\bibfnamefont{D.}~\bibnamefont{Xiao}},
  \bibinfo{author}{\bibfnamefont{M.-C.} \bibnamefont{Chang}}, \bibnamefont{and}
  \bibinfo{author}{\bibfnamefont{Q.}~\bibnamefont{Niu}}, \bibinfo{journal}{Rev.
  Mod. Phys.} \textbf{\bibinfo{volume}{82}}, \bibinfo{pages}{1959}
  (\bibinfo{year}{2010}),
  \urlprefix\url{https://link.aps.org/doi/10.1103/RevModPhys.82.1959}.

\bibitem[{\citenamefont{White et~al.}(1989)\citenamefont{White, Scalapino,
  Sugar, Loh, Gubernatis, and Scalettar}}]{white89}
\bibinfo{author}{\bibfnamefont{S.~R.} \bibnamefont{White}},
  \bibinfo{author}{\bibfnamefont{D.~J.} \bibnamefont{Scalapino}},
  \bibinfo{author}{\bibfnamefont{R.~L.} \bibnamefont{Sugar}},
  \bibinfo{author}{\bibfnamefont{E.~Y.} \bibnamefont{Loh}},
  \bibinfo{author}{\bibfnamefont{J.~E.} \bibnamefont{Gubernatis}},
  \bibnamefont{and} \bibinfo{author}{\bibfnamefont{R.~T.}
  \bibnamefont{Scalettar}}, \bibinfo{journal}{Phys. Rev. B}
  \textbf{\bibinfo{volume}{40}}, \bibinfo{pages}{506} (\bibinfo{year}{1989}),
  \urlprefix\url{https://link.aps.org/doi/10.1103/PhysRevB.40.506}.

\bibitem[{\citenamefont{Hirsch}(1985)}]{dqmc}
\bibinfo{author}{\bibfnamefont{J.~E.} \bibnamefont{Hirsch}},
  \bibinfo{journal}{Phys. Rev. B} \textbf{\bibinfo{volume}{31}},
  \bibinfo{pages}{4403} (\bibinfo{year}{1985}),
  \urlprefix\url{https://link.aps.org/doi/10.1103/PhysRevB.31.4403}.

\bibitem[{\citenamefont{White}(1992)}]{white92}
\bibinfo{author}{\bibfnamefont{S.~R.} \bibnamefont{White}},
  \bibinfo{journal}{Phys. Rev. Lett.} \textbf{\bibinfo{volume}{69}},
  \bibinfo{pages}{2863} (\bibinfo{year}{1992}),
  \urlprefix\url{https://link.aps.org/doi/10.1103/PhysRevLett.69.2863}.

\bibitem[{\citenamefont{Bauer et~al.}(2011)\citenamefont{Bauer, Carr, Evertz,
  Feiguin, Freire, Fuchs, Gamper, Gukelberger, Gull, Guertler
  et~al.}}]{Bauer2011}
\bibinfo{author}{\bibfnamefont{B.}~\bibnamefont{Bauer}},
  \bibinfo{author}{\bibfnamefont{L.~D.} \bibnamefont{Carr}},
  \bibinfo{author}{\bibfnamefont{H.~G.} \bibnamefont{Evertz}},
  \bibinfo{author}{\bibfnamefont{A.}~\bibnamefont{Feiguin}},
  \bibinfo{author}{\bibfnamefont{J.}~\bibnamefont{Freire}},
  \bibinfo{author}{\bibfnamefont{S.}~\bibnamefont{Fuchs}},
  \bibinfo{author}{\bibfnamefont{L.}~\bibnamefont{Gamper}},
  \bibinfo{author}{\bibfnamefont{J.}~\bibnamefont{Gukelberger}},
  \bibinfo{author}{\bibfnamefont{E.}~\bibnamefont{Gull}},
  \bibinfo{author}{\bibfnamefont{S.}~\bibnamefont{Guertler}},
  \bibnamefont{et~al.}, \bibinfo{journal}{Journal of Statistical Mechanics:
  Theory and Experiment} \textbf{\bibinfo{volume}{2011}},
  \bibinfo{pages}{P05001} (\bibinfo{year}{2011}),
  \urlprefix\url{https://doi.org/10.1088%2F1742-5468%2F2011%2F05%2Fp05001}.

\bibitem[{\citenamefont{Paiva and dos Santos}(2000)}]{thereza2000}
\bibinfo{author}{\bibfnamefont{T.}~\bibnamefont{Paiva}} \bibnamefont{and}
  \bibinfo{author}{\bibfnamefont{R.~R.} \bibnamefont{dos Santos}},
  \bibinfo{journal}{Phys. Rev. B} \textbf{\bibinfo{volume}{62}},
  \bibinfo{pages}{7007} (\bibinfo{year}{2000}),
  \urlprefix\url{https://link.aps.org/doi/10.1103/PhysRevB.62.7007}.

\end{thebibliography}


\end{document}